\documentclass[12pt]{article}
\usepackage{a4wide}
\usepackage{latexsym}
\usepackage{cite}
\usepackage{axodraw}
\usepackage{amssymb}

\usepackage{pslatex}
\usepackage[latin1]{inputenc}
\usepackage[T1]{fontenc}

\newcommand{\bq}{\begin{eqnarray}}
\newcommand{\eq}{\end{eqnarray}}
\renewcommand{\l}{\langle}
\renewcommand{\r}{\rangle} 
\newcommand{\eps}{\varepsilon}

\begin{document}

\thispagestyle{empty}

\begin{flushright}
  MZ-TH/05-05 \\
\end{flushright}

\vspace{1.5cm}

\begin{center}
  {\Large\bf Scalar diagrammatic rules for Born amplitudes in QCD\\
  }
  \vspace{1cm}
  {\large Christian Schwinn and Stefan Weinzierl\\
  \vspace{1cm}
      {\small \em Institut f{\"u}r Physik, Universit{\"a}t Mainz,}\\
      {\small \em D - 55099 Mainz, Germany}\\
  } 
\end{center}

\vspace{2cm}

\begin{abstract}\noindent
  {
We show that all Born amplitudes in QCD can be calculated
from scalar propagators and a set of three- and four-valent vertices.
In particular, our approach includes amplitudes with any number of quark pairs.
The quarks may be massless or massive.
The proof of the formalism is given entirely within quantum field theory.
   }
\end{abstract}

\vspace*{\fill}

\newpage

\section{Introduction}
\label{sect:intro}

The computation of QCD amplitudes is vital to extract information 
from ongoing and future collider experiments.
Even at Born level, the calculation of amplitudes with a large number of external legs is far
from trivial.
Efficient methods employed up to now start from a decomposition of QCD Born amplitudes into
gauge-independent colour 
structures \cite{Cvitanovic:1980bu,Berends:1987cv,Mangano:1987xk,Kosower:1987ic,Bern:1990ux}. 
The expression multiplying a specific colour structure
is usually called a partial amplitude.
The partial amplitudes can be calculated by using the 
spinor helicity formalism \cite{Berends:1981rb,DeCausmaecker:1982bg,Gunion:1985vc,Xu:1987xb,Gastmans:1990xh}
and by using recursive techniques \cite{Berends:1987me,Kosower:1989xy}.

It has been known for a long time, that helicity amplitudes for specific helicity combinations have a remarkably
simple analytic formula or vanish altogether \cite{Parke:1986gb,Berends:1987me}.
In  particular, the pure gluon amplitude vanishes if all gluons have the same helicity, or if all gluons except one
have the same helicity.
The first non-vanishing amplitude is obtained if $n-2$ gluon have one type of helicity, and $2$ gluons the other type.
The $n$-gluon amplitude with $n-2$ gluons of positive helicity and $2$ gluons of negative helicity is called a
maximal-helicity violating amplitude (MHV amplitude).
Similar, the $n$-gluon amplitude with $2$ gluons of positive helicity and $n-2$ gluons of negative helicity
is often referred to as $\overline{\mbox{MHV}}$-amplitude or ``googly'' amplitude.
For these amplitudes compact analytical formulae are known.

Recently,  Cachazo, Svr\v{c}ek and Witten \cite{Cachazo:2004kj}
proposed that the gluon amplitude with an
arbitrary helicity configuration can be calculated from diagrams with scalar propagators and new vertices,
which are MHV-amplitudes continued off-shell.
This conjecture was inspired by insight gained from the transformation of the amplitude to 
twistor space \cite{Witten:2003nn}.
In the latter publication, Witten conjectured that the $n$-gluon amplitude with $l$-loops is non-zero only if all
points lie in twistor space on an algebraic curve of degree $d$. The degree $d$ of this curve is given
by the number of negative helicity gluons plus the number of loops minus one.

These two conjectures triggered significant interest in the community and have been checked by explicit
calculations of Born amplitudes \cite{Roiban:2004vt,Roiban:2004ka,Roiban:2004yf,Gukov:2004ei,Kosower:2004yz,Zhu:2004kr}.
As the conjecture on twistor space makes a statement on loop amplitudes, even one-loop amplitudes have been studied 
extensively \cite{Cachazo:2004zb,Cachazo:2004by,Cachazo:2004dr,Britto:2004nj,Britto:2004tx,Britto:2004nc,Bena:2004xu,Bern:2004ky,Bidder:2004tx,Bidder:2004vx,Bern:2004bt,Bern:2005hs,Bidder:2005ri,Brandhuber:2004yw,Bedford:2004py,Bedford:2004nh,Quigley:2004pw,Roiban:2004ix,Luo:2004ss,Luo:2004nw}.
Within this context it is worth mentioning some earlier work on twistor methods used to simplify the calculation of Feynman diagrams
\cite{Chalmers:1997ui,Chalmers:1998jb,Chalmers:2001cy,Siegel:1999ew}.

Of practical interest for the calculation of Born amplitudes in QCD are recursive methods. 
These have been considered 
for the pure gluon case in \cite{Bena:2004ry,Britto:2004ap,Britto:2005fq}.
Furthermore, QCD amplitudes may involve in addition to gluons also quark pairs.
It has been known for a long time, that Born amplitudes with massless quarks may be related to pure gluon amplitudes through
super-symmetric relations \cite{Parke:1985pn,Grisaru:1977px}.
This is related to the fact that in partial amplitudes, where all colour information has been stripped off,
nothing distinguishes a (massless) quark from a gluino.
QCD amplitudes with massless quarks have been studied within the context of the MHV-approach in
\cite{Georgiou:2004wu,Georgiou:2004by,Khoze:2004ba,Wu:2004fb,Wu:2004jx,Su:2004ym,Luo:2005rx,Luo:2005my}.
Furthermore, the inclusion of additional non-QCD-type particles, like vector bosons or the Higgs boson
have been studied as well \cite{Bern:2004ba,Dixon:2004za,Badger:2004ty}.

In this paper we would like to extend the studies towards massive quarks.
We consider Born amplitudes in QCD with any number of quark pairs. The quarks may be massless or massive.
This allows for the application of our results towards processes involving the top quark.
With massive quarks we can no longer rely on super-symmetric relations 
and we develop therefore an alternative diagrammatic approach. 
We show that all Born amplitudes in QCD can be calculated from scalar propagators and a set
of three- and four-valent vertices.
We call these vertices ``primitive vertices'' to distinguish them 
from the vertices of the standard Feynman rules on the one hand,
and from the MHV-vertices on the other hand.
A helicity label $\{+,-\}$ is attached to each leg of a 
primitive vertex and propagators connect ``$+$''-labels with ``$-$''-labels.
One can define the degree of a vertex as the number of ``-''-labels minus one.
We show that only primitive vertices of degree zero and one occur.
The proof of the formalism is given entirely within quantum field theory.

The benefits of our approach are two-fold: First, we are able to treat all QCD partons on an equal footing, 
independently if they are massless or not. 
This allows the application of our approach towards problems involving massive quarks.
As all propagators are scalars no contraction of Lorentz- or spinor-indices is present any more.
This makes our method well suited for a 
fast implementation on a computer.

Secondly, we see the structure of the
Cachazo - Svr\v{c}ek - Witten prescription emerge from our diagrammatic approach: 
As the degree of an amplitude is exactly the sum of the degrees of the primitive vertices, 
a MHV amplitude is of degree one and contains exactly one primitive vertex of degree one, 
which is dressed up in all possible ways with degree zero vertices.
Similar, a pure gluon amplitude with three gluons of negative helicities is of degree two
and contains two vertices of degree
one, which again are combined in all possible ways with degree zero vertices.
In general, the pure gluon amplitude with $n_-$ gluons of negative helicity contains 
$(n_- - 1)$ primitive vertices of degree one, which are supplemented in all possible ways with degree zero vertices.

This paper is organised as follows:
In the next section we introduce our notation and review basic facts on QCD amplitudes.
In section \ref{sec:formalism} we show that Born amplitudes in QCD can be calculated from
scalar propagators and a set of vertices.
In section \ref{sect:discussion} we discuss the combinatorial aspects of our approach.
Section \ref{sect:concl} contains our conclusions.
In appendix \ref{appendix:spinors} we list useful information on spinors.
We summarise the scalar diagrammatic rules in appendix \ref{appendix:rules}.


\section{Notation and review of basic facts on QCD amplitudes}
\label{sect:notation}

In this paper we use the normalisation
\bq
 \mbox{Tr}\;T^a T^b & = & \frac{1}{2} \delta^{a b}
\eq
for the colour matrices.
Amplitudes in QCD may be decomposed into group-theoretical factors (carrying the colour structures)
multiplied by kinematic functions called partial amplitudes
\cite{Cvitanovic:1980bu,Berends:1987cv,Mangano:1987xk,Kosower:1987ic,Bern:1990ux}. 
These partial amplitudes do not contain any colour information and are gauge-invariant objects. 
\\
\\
The colour decomposition is obtained by replacing the structure constants $f^{abc}$
by
\bq
 i f^{abc} & = & 2 \left[ \mbox{Tr}\left(T^a T^b T^c\right) - \mbox{Tr}\left(T^b T^a T^c\right) \right] 
\eq
which follows from $ \left[ T^a, T^b \right] = i f^{abc} T^c$.
The resulting traces and strings of colour matrices can be further simplified with
the help of the Fierz identity :
\bq
 T^a_{ij} T^a_{kl} & = &  \frac{1}{2} \left( \delta_{il} \delta_{jk}
                         - \frac{1}{N} \delta_{ij} \delta_{kl} \right).
\eq
In the pure gluonic case tree level amplitudes with $n$ external gluons may be written in the form
\bq
{\cal A}_{n}(1,2,...,n) & = & g^{n-2} \sum\limits_{\sigma \in S_{n}/Z_{n}} 2 \; \mbox{Tr} \left(
 T^{a_{\sigma(1)}} ... T^{a_{\sigma(n)}} \right)
 A_{n}\left( \sigma(1), ..., \sigma(n) \right), 
\eq
where the sum is over all non-cyclic permutations of the external gluon legs.
The quantities $A_n(\sigma(1),...,\sigma(n))$, called the partial amplitudes, contain the kinematic information.
They are colour-ordered, e.g. only diagrams with a particular cyclic ordering of the gluons contribute.
\\
\\
The colour decomposition for a tree amplitude with a pair of quarks is
\bq
{\cal A}_{n+2}(q,1,2,...,n,\bar{q}) & = & g^{n} \sum\limits_{S_n} \left( T^{a_{\sigma(1)}} ... T^{a_{\sigma(n)}} \right)_{i_q j_{\bar{q}}}
A_{n+2}(q,\sigma(1),\sigma(2),...,\sigma(n),\bar{q}). 
\eq
where the sum is over all permutations of the gluon legs. 
Similar decompositions may be obtained for amplitudes with more than one pair of quarks.
\\
\\
The calculation of partial amplitudes in QCD is the main subject of this paper. In principle, they can be calculated
in a helicity basis from all Feynman diagrams with a fixed cyclic ordering.
The Feynman rules for colour-ordered partial amplitudes read:
\bq
\begin{picture}(100,35)(0,55)
\Vertex(50,50){2}
\Gluon(50,50)(50,80){3}{4}
\Gluon(50,50)(76,35){3}{4}
\Gluon(50,50)(24,35){3}{4}
\LongArrow(56,70)(56,80)
\LongArrow(67,47)(76,42)
\LongArrow(33,47)(24,42)
\Text(60,80)[lt]{$k_{1}$,$\mu$}
\Text(78,35)[lc]{$k_{2}$,$\nu$}
\Text(22,35)[rc]{$k_{3}$,$\rho$}
\end{picture}
 & = &
 i \left[
          g^{\mu\nu} \left( k_1^\rho - k_2^\rho \right)
        + g^{\nu\rho} \left( k_2^\mu - k_3^\mu \right)
        + g^{\rho\mu} \left( k_3^\nu - k_1^\nu \right)
   \right],
 \nonumber \\
\begin{picture}(100,75)(0,50)
\Vertex(50,50){2}
\Gluon(50,50)(71,71){3}{4}
\Gluon(50,50)(71,29){3}{4}
\Gluon(50,50)(29,29){3}{4}
\Gluon(50,50)(29,71){3}{4}
\Text(72,72)[lb]{$\mu$}
\Text(72,28)[lt]{$\nu$}
\Text(28,28)[rt]{$\rho$}
\Text(28,72)[rb]{$\sigma$}
\end{picture}
 & = &
   i \left[
        2 g^{\mu\rho} g^{\nu\sigma} - g^{\mu\nu} g^{\rho\sigma} 
                                     - g^{\mu\sigma} g^{\nu\rho}
 \right],
 \nonumber \\
\begin{picture}(100,75)(0,50)
\Vertex(50,50){2}
\Gluon(50,50)(80,50){3}{4}
\ArrowLine(29,29)(50,50)
\ArrowLine(50,50)(29,71)
\Text(82,50)[l]{$\mu$}
\end{picture}
 & = &
 i \gamma^\mu.
 \nonumber \\
\eq
For the calculation of 
helicity amplitudes \cite{Berends:1981rb,DeCausmaecker:1982bg,Gunion:1985vc,Xu:1987xb,Gastmans:1990xh}
one chooses for the 
spinors corresponding to external massless quarks two-component Weyl spinors
$\l p \pm |$ and $| p \pm \r$. We use here the bra-ket-notation. The relation with the notation using dotted and undotted indices
is as follows:
\bq
|p+\rangle = p_B,          & & \langle p+| = p_{\dot{A}}, \\
|p-\rangle = p^{\dot{B}},  & & \langle p-| = p^A. 
\eq
For the polarisation vectors of the external gluons one uses
\bq
\label{gluon_pol_onshell}
\eps_{\mu}^{+}(k,q) = \frac{\langle q-|\gamma_{\mu}|k-\rangle}{\sqrt{2} \langle q- | k + \rangle},
 & &
\eps_{\mu}^{-}(k,q) = \frac{\langle q+|\gamma_{\mu}|k+\rangle}{\sqrt{2} \langle k + | q - \rangle},
\eq
where $k$ is the momentum of the gluon and $q$ is an arbitrary light-like reference momentum.
It is possible to extend the helicity formalism to massive 
spinors \cite{Kleiss:1985yh,Berends:1985gf,Ballestrero:1995jn,Dittmaier:1998nn,vanderHeide:2000fx}.
We have collected all necessary details on spinors in appendix \ref{appendix:spinors}.
Our conventions are such that we always take all gluon momenta outgoing and 
the momenta of quarks to flow in the direction of the fermion lines.


\section{Scalar propagators and primitive vertices}
\label{sec:formalism}

In this section we show that QCD amplitudes can be calculated from
scalar propagators and a set of three- and four-valent vertices.
We start by recalling the off-shell continuation of spinors in
paragraph \ref{subsect:offshell}.
We then treat first the pure gluonic case in paragraph \ref{subsect:pure_gluons}.
The complete discussion carries over to quarks, which are dealt with in
paragraph \ref{subsect:quarks}.


\subsection{Off-shell continuation}
\label{subsect:offshell}

Let $q$ be a null-vector, which will be kept fixed throughout the discussion.
Using $q$, any massive vector $k$ can be written as 
a sum of two null-vectors $k^\flat$ and $q$ \cite{Kosower:2004yz}:
\bq
\label{offshellcont}
k & = & k^\flat + \frac{k^2}{2kq} q.
\eq
Obviously, if $k^2=0$, we have $k = k^\flat$.
Note further that $2kq = 2k^\flat q$.
Using eq. (\ref{offshellcont}) we may associate to any four-vector $k$
a massless four-vector $k^\flat$.
Using the projection onto $k^\flat$ we define the off-shell
continuation of Weyl spinors as
\bq
\label{offshellcontspinor}
 | k \pm \r & \rightarrow & | k^\flat \pm \r,
 \nonumber \\
 \l k \pm | & \rightarrow & \l k^\flat \pm |.
\eq
We are going to use the following abbreviations:
\bq
 & &
 \left\l i j \right\r = \left\l k_i^\flat - | k_j^\flat + \right\r,
  \;\;\;\;\;\;
 \left[ i j \right] = \left\l k_i^\flat + | k_j^\flat - \right\r,
 \nonumber \\
 & &
 \left\l i- \left| j \pm k \right| l- \right\r =
 \left\l k^\flat_i- \left| k\!\!\!/^\flat_j \pm k\!\!\!/^\flat_k \right| k^\flat_l- \right\r. 
\eq
Let us define an ``off-shell amplitude''
\bq
 O_n\left(1^{\lambda_1}, 2^{\lambda_2}, ..., n^{\lambda_n}\right),
\eq
depending on $n$ external momenta $k_i$ and helicities $\lambda_i$. The momenta need not be on-shell,
but momentum conservation is imposed:
\bq
 \sum\limits_{\mbox{\small gluons}} k_l 
 \;\;\; + \sum\limits_{\mbox{\small outgoing quarks}} k_i 
 \;\;\; - \sum\limits_{\mbox{\small incoming quarks}} k_j & = & 0.
\eq
By definition,
the off-shell amplitudes $O_n$ are calculated from all Feynman diagrams contributing to the
cyclic-ordered partial amplitude $A_n$, by using the off-shell
continuation eq. (\ref{offshellcontspinor}) 
for all external spinors and polarisation vectors, and by using the axial gauge
for all internal gluon propagators.
Compared to off-shell currents, which are used in recurrence 
relations of Berends-Giele type, an off-shell amplitude may have more 
than one leg off-shell.
By construction, if all external particles are on-shell, the off-shell
amplitude $O_n$ coincides with the physical amplitude $A_n$.
Let $n_-$ be the number of partons with negative helicity
of the off-shell amplitude $O_n$.
Then we define the degree $d$ of the off-shell amplitude $O_n$ 
as 
\bq
 d & = & n_- - 1.
\eq


\subsection{The pure gluonic case}
\label{subsect:pure_gluons}

We may continue the expressions for the gluon polarisation vectors in
eq. (\ref{gluon_pol_onshell}) to $k^2 \neq 0$, by using the projection onto $k^\flat$:
\bq
\label{pol_vec_continuation}
\eps_{\mu}^{+}(k,q) = \frac{\langle q-|\gamma_{\mu}|k^\flat-\rangle}{\sqrt{2} \langle q- | k^\flat + \rangle},
 & &
\eps_{\mu}^{-}(k,q) = \frac{\langle q+|\gamma_{\mu}|k^\flat+\rangle}{\sqrt{2} \langle k^\flat + | q - \rangle}.
\eq
The off-shell continued polarisation vectors satisfy:
\bq
\label{pol_prop}
 \eps_{\mu}^{\pm}(k,q) k^\mu = 0,
  \;\;\;\;\;
 \eps_{\mu}^{\pm}(k,q) q^\mu = 0,
  \;\;\;\;\;
 \eps_{\mu}^{+}(k_1,q) \eps^{+ \mu}(k_2,q) = 0,
  \;\;\;\;\;
 \eps_{\mu}^{-}(k_1,q) \eps^{- \mu}(k_2,q) = 0.
\eq
The polarisation sum is
\bq
\sum\limits_{\lambda = +/-} \varepsilon_\mu^{\lambda}(k,q) \varepsilon_\nu^{-\lambda}(k,q) 
 & = & 
 - g_{\mu \nu} + 2 \frac{k^\flat_\mu q_\nu + q_\mu k^\flat_\nu }{2 k q}.
\eq
The gluon propagator in the axial gauge is given by
\bq
\label{axial_gauge}
 \frac{i}{k^2} d_{\mu \nu} & = &
 \frac{i}{k^2} \left(  - g_{\mu \nu} + 2 \frac{k_\mu q_\nu + q_\mu k_\nu }{2 k q} \right).
\eq
If we introduce an unphysical polarisation 
\bq
 \eps_\mu^0(k,q) & = & 2 \frac{\sqrt{k^2}}{2kq} q_\mu,
\eq
the propagator can be written as
\bq
\label{basic_eq_gluon_prop}
 \frac{i}{k^2} d_{\mu \nu} & = &
 \frac{i}{k^2} 
 \left(
        \eps_\mu^+ \eps_\nu^-
      + \eps_\mu^- \eps_\nu^+
      + \eps_\mu^0 \eps_\nu^0
 \right).
\eq
This has already the structure of a scalar propagator $i/k^2$ times a sum over polarisations.
The additional polarisation $\eps^0$ can be absorbed into a redefinition of the
four-gluon vertex. This has already been noted in \cite{Kosower:1989xy}.
We define three- and four-valent primitive vertices as follows:
The three-valent vertex is just is just the three-gluon vertex contracted with
off-shell polarisation vectors:
\bq
 V_3(1^{\lambda_1},2^{\lambda_2},3^{\lambda_3})
 & = &
 O_3(1^{\lambda_1},2^{\lambda_2},3^{\lambda_3}).
\eq
From the properties of the polarisation vectors it follows, that $V_3(1^+,2^+,3^+)$ and $V_3(1^-,2^-,3^-)$
vanish. Non-zero values are obtained for 
\bq
 V_3(1^-,2^-,3^+) & = & 
 i \sqrt{2} \l 1 2 \r \frac{[3q]^2}{[1q][2q]}
 =
 i \sqrt{2} \frac{\l 1 2 \r^4}{\l 12 \r \l 23 \r \l 31 \r},
 \nonumber \\
 V_3(1^+,2^+,3^-) & = & 
 i \sqrt{2} [ 2 1 ] \frac{\l 3 q \r^2}{\l 1 q \r \l 2 q \r}
 =
 i \sqrt{2} \frac{[ 2 1 ]^4}{[ 32 ] [ 21 ] [ 13 ]},
\eq
and cyclic permutations thereof.
Let us now consider the polarisation $\eps^0_\mu$. The contraction of $\eps^0_\mu$
with $\eps^{+ \mu}$, $\eps^{- \mu}$ or $\eps^{0 \mu}$ vanishes.
Therefore any contraction of $\eps^0_\mu$ into a four-gluon vertex will give a vanishing contribution.
The only non-zero contribution is obtained from a contraction of a single $\eps^0$ into a three-gluon
vertex. In this case the other helicities are necessarily $\eps^+$ and $\eps^-$.
It follows that on both ends of a $\eps^0_\mu \eps^0_\nu$ propagator
there is always a three-gluon vertex attached.
Therefore we can absorb the $\eps^0$-polarisations into
a redefinition of the four-gluon vertex.
This leads to the definition of the
the primitive four-valent vertex, which is obtained 
from the sum of the standard four-gluon vertex plus contributions resulting
from two three-gluon vertices and an $\eps^0_\mu \eps^0_\nu$
propagator.
\begin{figure}
\begin{center}
\begin{tabular}{ccc}
\begin{picture}(110,100)(0,0)
 \Vertex(50,50){2}
 \Gluon(50,50)(80,80){4}{4}
 \Gluon(50,50)(80,20){4}{4}
 \Gluon(50,50)(20,20){4}{4}
 \Gluon(50,50)(20,80){4}{4}
 \Text(83,80)[l]{$1$}
 \Text(83,20)[l]{$2$}
 \Text(17,20)[r]{$3$}
 \Text(17,80)[r]{$4$}
\end{picture}
&
\begin{picture}(110,100)(0,0)
 \Vertex(70,50){2}
 \Vertex(30,50){2}
 \Gluon(30,50)(70,50){4}{4}
 \Gluon(70,50)(100,80){4}{4}
 \Gluon(70,50)(100,20){4}{4}
 \Gluon(30,50)(0,20){4}{4}
 \Gluon(30,50)(0,80){4}{4}
 \Text(103,80)[l]{$1$}
 \Text(103,20)[l]{$2$}
 \Text(-3,20)[r]{$3$}
 \Text(-3,80)[r]{$4$}
 \Text(35,57)[b]{$0$}
 \Text(65,57)[b]{$0$}
\end{picture}
&
\begin{picture}(110,100)(0,0)
 \Vertex(50,80){2}
 \Vertex(50,20){2}
 \Gluon(50,20)(50,80){-4}{5}
 \Gluon(50,80)(80,80){4}{3}
 \Gluon(50,20)(80,20){4}{3}
 \Gluon(50,20)(20,20){4}{3}
 \Gluon(50,80)(20,80){4}{3}
 \Text(83,80)[l]{$1$}
 \Text(83,20)[l]{$2$}
 \Text(17,20)[r]{$3$}
 \Text(17,80)[r]{$4$}
 \Text(56,30)[l]{$0$}
 \Text(56,70)[l]{$0$}
\end{picture}
\\
\end{tabular}
\caption{\label{fig:fourvalentvertex} Diagrams contributing to the four-valent gluon vertex.
In the last two diagrams only the $\eps^0$ polarisation is exchanged.}
\end{center}
\end{figure}
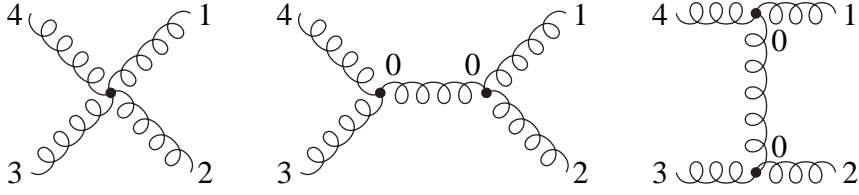
The contributions to the four-valent vertex are shown in fig. \ref{fig:fourvalentvertex}.
We find that the only non-zero contributions are
\bq
 V_4(1^+,2^+,3^-,4^-) & = &
 i \frac{[1q][2q]\l 3 q \r \l 4 q \r}{\l 1 q \r \l 2 q \r [3q] [4q]}
   \left( 
          1 + \frac{\left\l q- \left| 2-3 \right| q- \right\r \left\l q- \left| 4-1 \right| q- \right\r}
                   {\left\l q- \left| 2+3 \right| q- \right\r \left\l q- \left| 4+1 \right| q- \right\r} 
   \right),
 \nonumber \\
 V_4(1^+,2^-,3^+,4^-) & = & 
 i \frac{[1q] \l 2q \r [3q] \l 4 q \r}{\l 1 q \r [2q] \l 3 q \r [4q]}
   \left( 
          \frac{\left\l q- \left| 1-2 \right| q- \right\r \left\l q- \left| 3-4 \right| q- \right\r}
               {\left\l q- \left| 1+2 \right| q- \right\r \left\l q- \left| 3+4 \right| q- \right\r} 
 \right. \nonumber \\
 && \left.
        + \frac{\left\l q- \left| 2-3 \right| q- \right\r \left\l q- \left| 4-1 \right| q- \right\r}
               {\left\l q- \left| 2+3 \right| q- \right\r \left\l q- \left| 4+1 \right| q- \right\r} 
          - 2
   \right),
\eq
and permutations thereof.
We could in principle  eliminate the explicit dependence on the reference vector $q$, by noting that
\bq
 k_1^\flat + k_2^\flat + k_3^\flat + k_4^\flat 
 & = & 
 - \left( \frac{k_1^2}{2 k_1 q} + \frac{k_2^2}{2 k_2 q} + \frac{k_3^2}{2 k_3 q} + \frac{k_4^2}{2 k_4 q} \right) q.
\eq
For example, the prefactor becomes in this case 
\bq
 \frac{[1q] \l 2q \r [3q] \l 4 q \r}{\l 1 q \r [2q] \l 3 q \r [4q]}
 & = & 
 \frac{ \l 1+ | 2+3 | 4+ \r \l 2+ | 4+1 | 3+ \r}{ \l 1- | 2+3 | 4- \r \l 2- | 4+1 | 3- \r}.
\eq
Let $n_-$ be the number of particles with negative helicity of a vertex. As for off-shell amplitudes we define
the degree $d$ of a vertex as 
\bq
 d & = & n_- - 1.
\eq
We observe that for each vertex we have at least one parton of negative helicity, but never more than two 
parton with negative helicity.
Therefore all vertices are of degree zero or one.


\subsection{Quarks}
\label{subsect:quarks}

The off-shell continuation to $p^2 \neq m^2$ for spinors is straightforward:
\bq 
\label{quarkoffshellpol}
 u(-) = | p^\flat + \r + \frac{m}{\l p^\flat + | q - \r} | q - \r,
 & &
\bar{u}(+) = \l p^\flat + | + \frac{m}{\l q - | p^\flat + \r} \l q - |,
 \nonumber \\
 u(+) = | p^\flat - \r + \frac{m}{\l p^\flat - | q + \r} | q + \r,
 & &
\bar{u}(-) = \l p^\flat - | + \frac{m}{\l q + | p^\flat - \r} \l q + |.
\eq
These spinors have a smooth massless limit and the off-shell continuation of a massless quark
is simply
\bq 
 u(-) = | p^\flat + \r,
 & &
\bar{u}(+) = \l p^\flat + |,
 \nonumber \\
 u(+) = | p^\flat - \r,
 & &
\bar{u}(-) = \l p^\flat - |.
\eq
It is therefore sufficient to discuss the generic case of massive quarks.
Massless quarks are obtained by the $m\rightarrow 0$ limit.
The polarisation sum of the spinors in eq. (\ref{quarkoffshellpol}) reads
\bq
\sum\limits_{\lambda=+/-} u(-\lambda) \bar{u}(\lambda) 
 & = & p\!\!\!/ + m - \frac{p^2-m^2}{2pq} q\!\!\!\!\!/.
\eq
Therefore we can rewrite the quark propagator as
\bq
i \frac{p\!\!\!/ + m}{p^2-m^2}
 & = & 
 \frac{i}{p^2-m^2} 
 \left( \sum\limits_{\lambda=+/-} u(-\lambda) \bar{u}(\lambda) + \frac{p^2-m^2}{2pq} q\!\!\!\!\!/ \right).
\eq
We note the similarity with eq. (\ref{basic_eq_gluon_prop}). As the gluon propagator, the quark propagator
can be written as a polarisation sum of the off-shell continued polarisation vectors
and a piece proportional to $q\!\!\!\!\!/$. We say that this piece describes the exchange 
of an unphysical ``0''-polarisation.
Again, we can re-absorb the piece proportional to $q\!\!\!\!\!/$ into a four-valent vertex.
To do so, we have to take into account that also a gluon with an $\eps^0$-polarisation can couple
to a quark-gluon vertex.
It is easily verified, that for a quark-gluon vertex we can have no more than one
``0''-polarisation coupling to this vertex. (If we contract two or more ``0''-polarisations
into the vertex, we obtain zero.)
This ensures that on both ends of propagators carrying the ``0''-polarisation we have vertices, which contain
only the physical ``+/-'' polarisations. 
Therefore the effect of the exchange of ``0''-polarisations can be absorbed into the definition
of a four-valent vertex.
Let us now discuss the additional vertices due to the inclusion of quarks:
First, there is a three-valent vertex, obtained from the quark-gluon vertex 
contracted with the off-shell polarisation vectors.
\bq
 V_3(1_q^{\lambda_1},2_{\bar{q}}^{\lambda_2},3^{\lambda_3})
 & = &
 O_3(1_q^{\lambda_1},2_{\bar{q}}^{\lambda_2},3^{\lambda_3}).
\eq
Non-zero contributions are obtained for massless or massive quarks for the helicity configurations
\bq
 V_3(1_q^+,2_{\bar{q}}^-,3^+),
 \;\;\;
 V_3(1_q^+,2_{\bar{q}}^-,3^-),
 \;\;\;
 V_3(1_q^-,2_{\bar{q}}^+,3^+),
 \;\;\;
 V_3(1_q^-,2_{\bar{q}}^+,3^-).
\eq
For massive quarks there are two additional helicity configurations
\bq
 V_3(1_q^+,2_{\bar{q}}^+,3^-),
 \;\;\;
 V_3(1_q^-,2_{\bar{q}}^-,3^+),
\eq
which induce a spin-flip along the quark line. These vertices are proportional to the mass of the quark
and vanish in the massless limit.
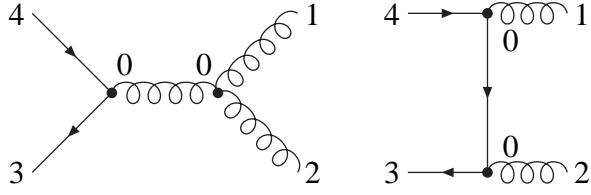
\begin{figure}
\begin{center}
\begin{tabular}{cc}
\begin{picture}(110,100)(0,0)
 \Vertex(70,50){2}
 \Vertex(30,50){2}
 \Gluon(30,50)(70,50){4}{4}
 \Gluon(70,50)(100,80){4}{4}
 \Gluon(70,50)(100,20){4}{4}
 \ArrowLine(30,50)(0,20)
 \ArrowLine(0,80)(30,50)
 \Text(103,80)[l]{$1$}
 \Text(103,20)[l]{$2$}
 \Text(-3,20)[r]{$3$}
 \Text(-3,80)[r]{$4$}
 \Text(35,57)[b]{$0$}
 \Text(65,57)[b]{$0$}
\end{picture}
&
\begin{picture}(110,100)(0,0)
 \Vertex(50,80){2}
 \Vertex(50,20){2}
 \ArrowLine(50,80)(50,20)
 \Gluon(50,80)(80,80){4}{3}
 \Gluon(50,20)(80,20){4}{3}
 \ArrowLine(50,20)(20,20)
 \ArrowLine(20,80)(50,80)
 \Text(83,80)[l]{$1$}
 \Text(83,20)[l]{$2$}
 \Text(17,20)[r]{$3$}
 \Text(17,80)[r]{$4$}
 \Text(56,30)[l]{$0$}
 \Text(56,70)[l]{$0$}
\end{picture}
\\
\end{tabular}
\caption{\label{fig:fourvalentquarkgluonvertex} Diagrams contributing to the four-valent quark-gluon vertex.
For the internal propagator only ``0''-polarisations are exchanged.
}
\end{center}
\end{figure}

The absorbtion of the unphysical ``0''-polarisations leads to two new four-valent vertices. First, there is a vertex
\bq
 V_4\left(1^{\lambda_1},2^{\lambda_2},3_q^{\lambda_3},4_{\bar{q}}^{\lambda_4}\right),
\eq
involving two quarks and two gluons. The Feynman diagrams contributing to this vertex are shown 
in fig. \ref{fig:fourvalentquarkgluonvertex}.
Non-zero contributions are obtained for the helicity configurations
\bq
 V_4\left(1^+,2^-,3_q^+,4_{\bar{q}}^-\right),
 \;\;\;
 V_4\left(1^-,2^+,3_q^+,4_{\bar{q}}^-\right),
 \;\;\;
 V_4\left(1^+,2^-,3_q^-,4_{\bar{q}}^+\right),
 \;\;\;
 V_4\left(1^-,2^+,3_q^-,4_{\bar{q}}^+\right).
\eq
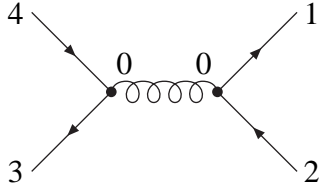
\begin{figure}
\begin{center}
\begin{tabular}{c}
\begin{picture}(110,100)(0,0)
 \Vertex(70,50){2}
 \Vertex(30,50){2}
 \Gluon(30,50)(70,50){4}{4}
 \ArrowLine(70,50)(100,80)
 \ArrowLine(100,20)(70,50)
 \ArrowLine(30,50)(0,20)
 \ArrowLine(0,80)(30,50)
 \Text(103,80)[l]{$1$}
 \Text(103,20)[l]{$2$}
 \Text(-3,20)[r]{$3$}
 \Text(-3,80)[r]{$4$}
 \Text(35,57)[b]{$0$}
 \Text(65,57)[b]{$0$}
\end{picture}
\\
\end{tabular}
\caption{\label{fig:fourvalentquarkvertex} Diagrams contributing to the four-valent quark-quark vertex.
For the internal propagator only ``0''-polarisations are exchanged.
}
\end{center}
\end{figure}
Secondly, there is a vertex
\bq
 V_4\left(1_q^{\lambda_1},2_{\bar{q}}^{\lambda_2},3_q^{\lambda_3},4_{\bar{q}}^{\lambda_4}\right),
\eq
involving two pairs of quarks. The Feynman diagram contributing to this vertex is shown 
in fig. \ref{fig:fourvalentquarkvertex}.
Non-zero contributions are obtained for the helicity configurations
\bq
 V_4\left(1_q^+,2_{\bar{q}}^-,3_q^+,4_{\bar{q}}^-\right),
 \;\;\;
  V_4\left(1_q^+,2_{\bar{q}}^-,3_q^-,4_{\bar{q}}^+\right).
\eq
We have summarised the formulae for all non-vanishing vertices in appendix \ref{appendix:rules}.
As in the pure gluon case, all vertices are of degree zero or one.


\section{Discussion}
\label{sect:discussion}

In the previous section we have shown, that all QCD amplitudes can be calculated from a set of three- or
four-valent vertices and scalar propagators.
We have defined the degree of a vertex as the number of negative helicities minus one.
As each vertex has either one or two negative helicities, we deal only with vertices of degree zero or one.
Further, one observes that all vertices of degree zero are three-valent, whereas vertices of degree one can
either be three-valent or four-valent.
We further defined the degree of an off-shell amplitude as the number of the external particles with negative
helicity minus one. As an off-shell amplitude is a sum of diagrams, each diagram inherits the degree of the off-shell
amplitude.
We would like to show that the degree of a diagram is the sum of the degrees of its vertices.
To this aim let $v_3^+$ be the number of three-valent vertices of degree zero in the diagram.
Similar, let $v_3^-$ be the number of three-valent vertices of degree one and
let $v_4^-$ be the number of four-valent vertices of degree one.
(Note that there are no four-valent vertices of degree zero.)
Since we are considering Born diagrams, the number of internal propagators $n_{prop}$ is just the number of vertices
minus one:
\bq
 n_{prop} & = & v_3^+ + v_3^- + v_4^- - 1.
\eq
In total we have
\bq
 v_3^+ + 2 v_3^- + 2 v_4^-
\eq
negative helicity labels in the diagram. As a propagator connects ``+''-helicities with ``-''-helicities, 
$n_{prop}$ negative helicity labels are attached to internal propagators.
Therefore the number $n_-$ of external particles with negative helicity is given by
\bq
 n_- & = & v_3^- + v_4^- + 1.
\eq
Therefore the degree $d$ of the diagram is given by
\bq
 d & = & n_- - 1 = v_3^- + v_4^-,
\eq
which shows that the degree $d$ of a diagram is the sum of the degrees of its vertices.

Let us consider the pure gluon off-shell amplitude $O_n(1^+, 2^+, ..., n^+)$, where all gluons have
positive helicity.
This amplitude would have degree $(-1)$. Since there are no vertices of negative degree, this off-shell
amplitude must vanish.
This actually proves that the first Parke-Taylor formula still holds if all particles are continued off-shell.

Of particular importance are off-shell amplitude of degree zero and one. From the argument above it follows that
off-shell amplitudes of degree zero must contain only vertices of degree zero.
In the pure gluonic case, there is only one degree zero vertex. As a consequence, the pure gluonic off-shell 
amplitude $O_n(1^+, 2^+, ..., (n-1)^+, n^-)$ of degree zero fulfills 
a rather simple recurrence relation:
\bq
\lefteqn{
 O_n\left( 1^+, ..., (n-1)^+, n^- \right)
 =  
 \sum\limits_{j=2}^{n-1} V_3( (-k_{1,j})^+, (-k_{j,n})^+,n^-)
} & &
 \nonumber \\
 & &
 \times 
 \frac{i}{k_{1,j}^2} O_j(1^+,...,(j-1)^+, k_{1,j}^- )
 \frac{i}{k_{j,n}^2} O_{n-j+1}(j^+,...,(n-1)^+, k_{j,n}^- ),
\eq
where
\bq 
 k_{i,j} & = & \sum\limits_{l=i}^{j-1} k_l,
\eq
and where we define the two-point amplitude to be the inverse propagator:
\bq
 O_2((j-1)^+, k_{(j-1),j}^- ) & = & - i k_{(j-1),j}^2.
\eq 
Off-shell amplitudes of degree one contain exactly one vertex of degree one. All other vertices are of degree
zero.
Since by definition an off-shell amplitude is just a sum of diagrams build out of the basic vertices in
all possible ways compatible with the cyclic ordering, an off-shell amplitude of degree one consists therefore
of diagrams containing a vertex of degree one dressed up in all possible ways compatible with the cyclic ordering
with vertices of degree zero.

Let us now consider an off-shell amplitude with $n_-$ external particles of negative helicity.
This amplitude is of degree $d=n_--1$ and each diagram has exactly $(n_--1)$ vertices of degree one.
The off-shell amplitude is just the sum of all diagrams, obtained from combining 
in all possible ways compatible with the cyclic ordering 
$(n_--1)$ vertices of degree one with the remaining vertices of degree zero.

We would like to comment on amplitudes with massive quarks.
Compared to the massless case, two new features make their appearance:
First there are helicity configurations, which vanish in the simultaneous massless 
and on-shell limit, but remain non-zero in the on-shell limit for non-zero masses.
Examples of this kind are amplitudes of the form
\bq
\label{onshellforbidden}
 A_n\left(1_Q^+, 2_{\bar{Q}}^-, 3^+, ..., n^+ \right).
\eq
Secondly, there are the helicity flip vertices
\bq
 V_3(1_q^+,2_{\bar{q}}^+,3^-),
 \;\;\;
 V_3(1_q^-,2_{\bar{q}}^-,3^+),
\eq
which are proportional to the mass and vanish therefore in the massless limit.
We note that $V_3(1_q^+,2_{\bar{q}}^+,3^-)$ is a degree zero vertex, whereas
$V_3(1_q^-,2_{\bar{q}}^-,3^+)$ has degree one.
Let us consider an amplitude with one massive quark pair. We first consider the helicity
configuration $Q^+ \bar{Q}^-$. Along the massive quark line we must have an equal number
of helicity flips induced by
$V_3(1_q^+,2_{\bar{q}}^+,3^-)$ and $V_3(1_q^-,2_{\bar{q}}^-,3^+)$.
Since the latter vertex is of degree one, the total number of helicty flips $f$
is bounded by the degree $d$ of the amplitude:
\bq
 f & \le & 2 d.
\eq
As an example it follows immediately 
from this bound that amplitudes of the form (\ref{onshellforbidden})
cannot contain a helicity flip.  
A similar argument applies to the helicity configuration $Q^+ \bar{Q}^+$.  
Here we have along the fermion line at least one helicity flip of degree zero 
together with at most $d$ additional pairs of degree one and zero helicity flips.
So the total number is bounded by 
\bq 
 f & \le & 2 d + 1.  
\eq 
For the helicity configuration $Q^- \bar{Q}^-$ there is at least one flip of
degree one, leaving at most $2(d-1)$ additional helicity flips so the
total number is bounded by 
\bq 
 f & \le & 2 d - 1.  
\eq


\section{Conclusions}
\label{sect:concl}

In this paper we considered Born amplitudes in QCD. Decomposing the full amplitude into gauge-invariant
partial amplitudes, we showed that the partial amplitudes can be calculated from simple rules,
involving only scalar propagators and a set of complex valued three- and four-valent vertices.
As no contraction of Lorentz- or spinor-indices is present any more, our method is well suited for a 
fast implementation on a computer.
Our approach is not restricted to gluons only, but treated gluons, massless quarks and massive quarks
on almost equal footing.
It is possible to assign to each vertex a degree, given by the number of negative helicities minus one.
It is a striking feature, that only vertices of degree zero and one occur.

\subsection*{Acknowledgements}

The work of C.S. has been supported
by the Deutsche Forschungsgemeinschaft through the
Graduiertenkolleg `Eichtheorien' at Mainz University.


\begin{appendix}

\section{Spinors}
\label{appendix:spinors}

For the metric we use
\bq
g_{\mu \nu} & = & \mbox{diag}(+1,-1,-1,-1).
\eq
We define the light-cone coordinates as
\bq
p_+ = p_0 + p_3, \;\;\; p_- = p_0 - p_3, \;\;\; p_{\bot} = p_1 + i p_2,
                                         \;\;\; p_{\bot^\ast} = p_1 - i p_2.
\eq
In terms of the light-cone components of a null-vector, the corresponding massless spinors $\l p \pm |$ and $| p \pm \r$ 
can be chosen as
\bq
\left| p+ \right\r = \frac{e^{-i \frac{\phi}{2}}}{\sqrt{\left| p_+ \right|}} \left( \begin{array}{c}
  -p_{\bot^\ast} \\ p_+ \end{array} \right),
 & &
\left| p- \right\r = \frac{e^{-i \frac{\phi}{2}}}{\sqrt{\left| p_+ \right|}} \left( \begin{array}{c}
  p_+ \\ p_\bot \end{array} \right),
 \nonumber \\
\left\l p+ \right| = \frac{e^{-i \frac{\phi}{2}}}{\sqrt{\left| p_+ \right|}} 
 \left( -p_\bot, p_+ \right),
 & &
\left\l p- \right| = \frac{e^{-i \frac{\phi}{2}}}{\sqrt{\left| p_+ \right|}} 
 \left( p_+, p_{\bot^\ast} \right),
\eq
where the phase $\phi$ is given by
\bq
p_+ & = & \left| p_+ \right| e^{i\phi_+}.
\eq
With these phase conventions we have the following relations between a spinor corresponding to a vector $p$ and a spinor
corresponding to a vector $(-p)$:
\bq
 \left| \left(-p\right) \pm \right\r & = & i \left| p \pm \right\r,
 \nonumber \\
 \left\l \left(-p\right) \pm \right| & = & i \left\l p \pm \right|.
\eq
Consequently, we have for the polarisation vectors of the gluon:
\bq
 \eps_\mu^\pm(-k,q) & = & \eps_\mu^\pm(k,q).
\eq
For the quarks we take the momentum to flow always along the fermion
line, i.e. we only consider in- and outgoing quarks with spinors $u$
and $\bar{u}$.
In- and outgoing anti-quarks with spinors $v$ and $\bar{v}$ are
related to the spinors $u$ and $\bar{u}$ by taking the momenta $p$
to $-p$ and a flip in helicity.
For massive quarks we have the spinors
\bq 
 u(+) & = & \frac{1}{\l p^\flat - | q + \r} \left( p\!\!\!/ + m \right) | q + \r
          = | p^\flat - \r + \frac{m}{\l p^\flat - | q + \r} | q + \r,
 \nonumber \\
 u(-) & = & \frac{1}{\l p^\flat + | q - \r} \left( p\!\!\!/ + m \right) | q - \r
          = | p^\flat + \r + \frac{m}{\l p^\flat + | q - \r} | q - \r.
\eq
For the conjugate spinors we have
\bq
\bar{u}(+) & = & \frac{1}{\l q - | p^\flat + \r} \l q - | \left( p\!\!\!/ + m \right)
             = \l p^\flat + | + \frac{m}{\l q - | p^\flat + \r} \l q - |,
 \nonumber \\
\bar{u}(-) & = & \frac{1}{\l q + | p^\flat - \r} \l q + | \left( p\!\!\!/ + m \right)
             = \l p^\flat - | + \frac{m}{\ q + | p^\flat - \r} \l q + |.
\eq
These spinors satisfy the Dirac equations
\bq 
\left( p\!\!\!/ - m \right) u(\lambda) = 0, & & 
\bar{u}(\lambda) \left( p\!\!\!/ - m \right) = 0,
\eq
the orthogonality relations
\bq
\bar{u}(\bar{\lambda}) u(-\lambda) & = & 2 m \delta_{\bar{\lambda}\lambda}, 
\eq
and the completeness relation
\bq
\sum\limits_{\lambda=+/-} u(-\lambda) \bar{u}(\lambda) & = & p\!\!\!/ + m.
\eq
Note that these spinors have a smooth massless limit. For $m=0$ we
obtain
\bq
 u(\pm) = | p \mp \r, 
 & &
 \bar{u}(\pm) = \l p \pm |.
\eq


\section{Summary of the diagrammatic rules}
\label{appendix:rules}

In this section we summarise the diagrammatic rules. Propagators are
always scalars, e.g.
\bq
 \frac{i}{p^2}
\eq
for gluons and massless quarks, and
\bq
 \frac{i}{p^2 - m^2}
\eq
for massive quarks.
They connect ``+''-labels with ``-''-labels.
The expressions for the vertices involve spinor products.
We recall our notation:
\bq
 & &
 \left\l i j \right\r = \left\l k_i^\flat - | k_j^\flat + \right\r,
  \;\;\;\;\;\;
 \left[ i j \right] = \left\l k_i^\flat + | k_j^\flat - \right\r,
 \nonumber \\
 & &
 \left\l i- \left| j \pm k \right| l- \right\r =
 \left\l k^\flat_i- \left| k\!\!\!/^\flat_j \pm k\!\!\!/^\flat_k \right| k^\flat_l- \right\r,
\eq
where $k^\flat$ is the projected light-like four-vector associated to $k$. 
$k^\flat$ is obtained from $k$ and the reference momentum $q$ as follows:
\bq
k^\flat & = & k - \frac{k^2}{2kq} q.
\eq
Our conventions for the momenta are such that we always take all gluon momenta outgoing and 
the momenta of quarks to flow in the direction of the fermion lines.
The vertices involving only gluons are:
\bq
\begin{picture}(100,35)(0,55)
\Vertex(50,50){2}
\Gluon(50,50)(50,80){3}{4}
\Gluon(50,50)(76,35){3}{4}
\Gluon(50,50)(24,35){3}{4}
\Text(60,80)[lt]{$1^-$}
\Text(78,35)[lc]{$2^-$}
\Text(22,35)[rc]{$3^+$}
\end{picture}
 & = &
 i \sqrt{2} \l 1 2 \r \frac{[3q]^2}{[1q][2q]}
 =
 i \sqrt{2} \frac{\l 1 2 \r^4}{\l 12 \r \l 23 \r \l 31 \r},
 \nonumber \\
\begin{picture}(100,75)(0,50)
\Vertex(50,50){2}
\Gluon(50,50)(50,80){3}{4}
\Gluon(50,50)(76,35){3}{4}
\Gluon(50,50)(24,35){3}{4}
\Text(60,80)[lt]{$1^+$}
\Text(78,35)[lc]{$2^+$}
\Text(22,35)[rc]{$3^-$}
\end{picture}
 & = &
 i \sqrt{2} [ 2 1 ] \frac{\l 3 q \r^2}{\l 1 q \r \l 2 q \r}
 =
 i \sqrt{2} \frac{[ 2 1 ]^4}{[ 32 ] [ 21 ] [ 13 ]},
 \nonumber \\
\begin{picture}(100,75)(0,50)
\Vertex(50,50){2}
\Gluon(50,50)(71,71){3}{4}
\Gluon(50,50)(71,29){3}{4}
\Gluon(50,50)(29,29){3}{4}
\Gluon(50,50)(29,71){3}{4}
\Text(72,72)[lb]{$1^+$}
\Text(72,28)[lt]{$2^+$}
\Text(28,28)[rt]{$3^-$}
\Text(28,72)[rb]{$4^-$}
\end{picture}
 & = &
 i \frac{[1q][2q]\l 3 q \r \l 4 q \r}{\l 1 q \r \l 2 q \r [3q] [4q]}
   \left( 
          1 + \frac{\left\l q- \left| 2-3 \right| q- \right\r \left\l q- \left| 4-1 \right| q- \right\r}
                   {\left\l q- \left| 2+3 \right| q- \right\r \left\l q- \left| 4+1 \right| q- \right\r} 
   \right),
 \nonumber \\
\begin{picture}(100,75)(0,50)
\Vertex(50,50){2}
\Gluon(50,50)(71,71){3}{4}
\Gluon(50,50)(71,29){3}{4}
\Gluon(50,50)(29,29){3}{4}
\Gluon(50,50)(29,71){3}{4}
\Text(72,72)[lb]{$1^+$}
\Text(72,28)[lt]{$2^-$}
\Text(28,28)[rt]{$3^+$}
\Text(28,72)[rb]{$4^-$}
\end{picture}
 & = &
 i \frac{[1q] \l 2q \r [3q] \l 4 q \r}{\l 1 q \r [2q] \l 3 q \r [4q]}
   \left( 
          \frac{\left\l q- \left| 1-2 \right| q- \right\r \left\l q- \left| 3-4 \right| q- \right\r}
               {\left\l q- \left| 1+2 \right| q- \right\r \left\l q- \left| 3+4 \right| q- \right\r} 
 \right. \nonumber \\
 && \left.
        + \frac{\left\l q- \left| 2-3 \right| q- \right\r \left\l q- \left| 4-1 \right| q- \right\r}
               {\left\l q- \left| 2+3 \right| q- \right\r \left\l q- \left| 4+1 \right| q- \right\r} 
          - 2
   \right).
 \nonumber \\
\eq
Three-valent vertices involving a pair of quarks are:
\bq
\begin{picture}(100,35)(0,55)
\Vertex(50,50){2}
\Gluon(50,50)(20,50){3}{4}
\ArrowLine(50,50)(71,71)
\ArrowLine(71,29)(50,50)
\Text(72,72)[lb]{$1^+$}
\Text(72,28)[lt]{$2^-$}
\Text(18,50)[r]{$3^+$}
\end{picture}
 & = &
 i \sqrt{2} [ 1 3 ] \frac{\l 2 q \r}{\l 3 q \r}
 =
 i \sqrt{2} \frac{[ 1 3 ]^2}{[ 1 2 ]},
 \nonumber \\
\begin{picture}(100,75)(0,50)
\Vertex(50,50){2}
\Gluon(50,50)(20,50){3}{4}
\ArrowLine(50,50)(71,71)
\ArrowLine(71,29)(50,50)
\Text(72,72)[lb]{$1^+$}
\Text(72,28)[lt]{$2^-$}
\Text(18,50)[r]{$3^-$}
\end{picture}
 & = &
 i \sqrt{2} \l 3 2 \r \frac{[ 1 q ]}{[ 3 q ]}
 =
 i \sqrt{2} \frac{\l 2 3 \r^2}{\l 2 1 \r},
 \nonumber \\
\begin{picture}(100,75)(0,50)
\Vertex(50,50){2}
\Gluon(50,50)(20,50){3}{4}
\ArrowLine(50,50)(71,71)
\ArrowLine(71,29)(50,50)
\Text(72,72)[lb]{$1^-$}
\Text(72,28)[lt]{$2^+$}
\Text(18,50)[r]{$3^+$}
\end{picture}
 & = &
 i \sqrt{2} [ 2 3 ] \frac{\l 1 q \r}{\l 3 q \r}
 =
 i \sqrt{2} \frac{[ 2 3 ]^2}{[ 1 2]},
 \nonumber \\
\begin{picture}(100,75)(0,50)
\Vertex(50,50){2}
\Gluon(50,50)(20,50){3}{4}
\ArrowLine(50,50)(71,71)
\ArrowLine(71,29)(50,50)
\Text(72,72)[lb]{$1^-$}
\Text(72,28)[lt]{$2^+$}
\Text(18,50)[r]{$3^-$}
\end{picture}
 & = &
 i \sqrt{2} \l 3 1 \r \frac{[ 2 q ]}{[ 3 q ]}
 =
 i \sqrt{2} \frac{\l 3 1 \r^2}{\l 2 1 \r},
 \nonumber \\
\begin{picture}(100,75)(0,50)
\Vertex(50,50){2}
\Gluon(50,50)(20,50){3}{4}
\ArrowLine(50,50)(71,71)
\ArrowLine(71,29)(50,50)
\Text(72,72)[lb]{$1^+$}
\Text(72,28)[lt]{$2^+$}
\Text(18,50)[r]{$3^-$}
\end{picture}
 & = &
 -i \sqrt{2} m \frac{\l 3 q \r^2}{\l 1 q \r \l 2 q \r}
 =
 i \sqrt{2} m \frac{[ 1 2]^2}{[ 2 3 ][ 3 1 ]},
 \nonumber \\
\begin{picture}(100,75)(0,50)
\Vertex(50,50){2}
\Gluon(50,50)(20,50){3}{4}
\ArrowLine(50,50)(71,71)
\ArrowLine(71,29)(50,50)
\Text(72,72)[lb]{$1^-$}
\Text(72,28)[lt]{$2^-$}
\Text(18,50)[r]{$3^+$}
\end{picture}
 & = &
 i \sqrt{2} m \frac{[ 3 q ]^2}{[ 1 q ] [ 2 q ]}
 =
 - i \sqrt{2} m \frac{\l 1 2 \r^2}{\l 2 3 \r \l 3 1 \r}.
 \nonumber \\
\eq
Note that the last two vertices involve a helicity flip along the fermion line and vanish in the massless limit.
The four-valent vertices with a pair of quarks and two gluons are:
\bq
\begin{picture}(100,35)(0,55)
\Vertex(50,50){2}
\Gluon(50,50)(71,71){3}{4}
\Gluon(50,50)(71,29){3}{4}
\ArrowLine(50,50)(29,29)
\ArrowLine(29,71)(50,50)
\Text(72,72)[lb]{$1^+$}
\Text(72,28)[lt]{$2^-$}
\Text(28,28)[rt]{$3^+$}
\Text(28,72)[rb]{$4^-$}
\end{picture}
 & = &
 2 i \frac{[ 1 q ] \l 2 q \r}{\l 1 q \r [ 2 q ]}
     \frac{[ 3 q ] \l 4 q \r}{\l q- | 3-4 | q- \r}
     \left( \frac{\l q- | 1-2 | q- \r}{\l q- | 1+2 | q- \r}
           +\frac{\l q- | 3-4 | q- \r}{\l q- | 2+3 | q- \r}
     \right),
 \nonumber \\
\begin{picture}(100,75)(0,50)
\Vertex(50,50){2}
\Gluon(50,50)(71,71){3}{4}
\Gluon(50,50)(71,29){3}{4}
\ArrowLine(50,50)(29,29)
\ArrowLine(29,71)(50,50)
\Text(72,72)[lb]{$1^-$}
\Text(72,28)[lt]{$2^+$}
\Text(28,28)[rt]{$3^+$}
\Text(28,72)[rb]{$4^-$}
\end{picture}
 & = &
 2 i \frac{\l 1 q \r [ 2 q ]}{[ 1 q ] \l 2 q \r}
     \frac{[ 3 q ] \l 4 q \r}{\l q- | 3-4 | q- \r}
     \frac{\l q- | 1-2 | q- \r}{\l q- | 1+2 | q- \r},
 \nonumber \\
\begin{picture}(100,75)(0,50)
\Vertex(50,50){2}
\Gluon(50,50)(71,71){3}{4}
\Gluon(50,50)(71,29){3}{4}
\ArrowLine(50,50)(29,29)
\ArrowLine(29,71)(50,50)
\Text(72,72)[lb]{$1^+$}
\Text(72,28)[lt]{$2^-$}
\Text(28,28)[rt]{$3^-$}
\Text(28,72)[rb]{$4^+$}
\end{picture}
 & = &
 2 i \frac{[ 1 q ] \l 2 q \r}{\l 1 q \r [ 2 q ]}
     \frac{\l 3 q \r [ 4 q ]}{\l q- | 3-4 | q- \r}
     \frac{\l q- | 1-2 | q- \r}{\l q- | 1+2 | q- \r},
 \nonumber \\
\begin{picture}(100,75)(0,50)
\Vertex(50,50){2}
\Gluon(50,50)(71,71){3}{4}
\Gluon(50,50)(71,29){3}{4}
\ArrowLine(50,50)(29,29)
\ArrowLine(29,71)(50,50)
\Text(72,72)[lb]{$1^-$}
\Text(72,28)[lt]{$2^+$}
\Text(28,28)[rt]{$3^-$}
\Text(28,72)[rb]{$4^+$}
\end{picture}
 & = &
 2 i \frac{\l 1 q \r [ 2 q ]}{[ 1 q ] \l 2 q \r}
     \frac{\l 3 q \r [ 4 q ]}{\l q- | 3-4 | q- \r}
     \left( \frac{\l q- | 1-2 | q- \r}{\l q- | 1+2 | q- \r}
           +\frac{\l q- | 3-4 | q- \r}{\l q- | 2+3 | q- \r}
     \right).
 \nonumber \\
\eq
Finally, we have the four-valent vertices with two pairs of quarks:
\bq
\begin{picture}(100,35)(0,55)
\ArrowLine(52,50)(73,71)
\ArrowLine(73,29)(52,50)
\ArrowLine(48,50)(27,29)
\ArrowLine(27,71)(48,50)
\Text(74,72)[lb]{$1^+$}
\Text(74,28)[lt]{$2^-$}
\Text(26,28)[rt]{$3^+$}
\Text(26,72)[rb]{$4^-$}
\end{picture}
 & = &
 4 i \frac{[ 1 q ] \l 2 q \r [ 3 q ] \l 4 q \r}{\l q- | 1-2 | q- \r \l q- | 3-4 | q- \r},
 \nonumber \\
\begin{picture}(100,75)(0,50)
\ArrowLine(52,50)(73,71)
\ArrowLine(73,29)(52,50)
\ArrowLine(48,50)(27,29)
\ArrowLine(27,71)(48,50)
\Text(74,72)[lb]{$1^+$}
\Text(74,28)[lt]{$2^-$}
\Text(26,28)[rt]{$3^-$}
\Text(26,72)[rb]{$4^+$}
\end{picture}
 & = &
 4 i \frac{[ 1 q ] \l 2 q \r \l 3 q \r [ 4 q ]}{\l q- | 1-2 | q- \r \l q- | 3-4 | q- \r}.
 \nonumber \\
\eq

\end{appendix}



\end{document}